\documentclass[12pt]{article}
\usepackage{epsf}
\usepackage{amsmath}
\usepackage{graphics}
\usepackage{cite}
\renewcommand{\thefootnote}{\#\arabic{footnote}}
\begin{document}

\newcommand{\gtrsim}{ \mathop{}_{\textstyle \sim}^{\textstyle >} }
\newcommand{\lesssim}{ \mathop{}_{\textstyle \sim}^{\textstyle <} }

\newcommand{\rem}[1]{{\bf #1}}

\renewcommand{\thefootnote}{\fnsymbol{footnote}}
\setcounter{footnote}{0}
\begin{titlepage}

\def\thefootnote{\fnsymbol{footnote}}

\vskip .5in
\bigskip
\bigskip
\begin{center}

{\bf \Large Dark Matter as Ultralight Axion-Like particle in $E_6 \times U(1)_X$ GUT
with QCD Axion \\}
\vspace{2.5cm}
{\bf Claudio Corian\`{o} and Paul H. Frampton \\  }
\vspace{0.5cm}
{Dipartimento di Matematica e Fisica "Ennio De Giorgi",\\ 
Universit\`{a} del Salento and INFN-Lecce,\\ Via Arnesano, 73100 Lecce, Italy\footnote{claudio.coriano@le.infn.it, paul.h.frampton@gmail.com}
}

\bigskip

\begin{abstract}
Axion-like fields are naturally generated by a mechanism of anomaly cancellation of one or more anomalous gauge abelian symmetries at the Planck scale, emerging as duals of a two-form from the massless bosonic sector of string theory. This suggests an analogy of the Green-Schwarz mechanism of anomaly cancellation, at field theory level, which results in one or more Stueckelberg pseudoscalars. In the case of a single Stueckelberg pseudoscalar $b$, vacuum misalignments at phase transitions in the early Universe at the GUT scale provide a small mass - due to instanton 
suppression of the periodic potential - for a component of $b$, denoted as $\chi$ and termed the "axi-Higgs", which is a physical axion-like particle.  The coupling of the axi-Higgs to the gauge sector via Wess-Zumino terms is suppressed by the Planck mass, which guarantees its decoupling, while its angle of misalignment is related to $M_{GUT}$. We build a gauged $E_6\times U(1)$ model with anomalous $U(1)$. It contains both an
automatic invisible QCD axion and an ultra-light axi-Higgs. The invisible axion present in the model solves the strong CP problem and has mass in the  conventional range while
the axi-Higgs, which can act as dark matter, is sufficiently light ($10^{-22} \textrm{ eV} < m_{\chi} <  10^{-20}  \textrm{ eV}$)  to solve short-distance problems which confront other cold dark matter candidates.
\end{abstract}

\end{center}

\end{titlepage}

\newpage
\section{Introduction} 
Axions have played a key role in offering a possible solution of the strong CP problem 
(see \cite{Kim:2008hd, ARingwald} for an overview) and searches for such particles have been pursued for a long time by several experimental groups, so far with a negative outcome. Two distinctive features of an axion solution - as derived from the original Peccei-Quinn (PQ) proposal \cite{Peccei:1977ur} and its extensions 
\cite{Kim:1979if,Shifman:1979if, Zhitnitsky:1980tq, Dine:1981rt}- are: \\
1) the appearance of a single scale $f_a$ which controls both their mass  and their coupling to the gauge fields, via an 
$a(x) F\tilde{F}$ operator, where $a(x)$ is the axion field and 2) their non-thermal decoupling at the hadron phase transition, characterised by a mechanism of vacuum misalignment. The latter causes axions to be a component of cold rather than hot dark matter, even for small values of their mass, currently expected to be in the $\mu$eV-meV range. 

The tight experimental constraints derived from the mass/coupling (single scale) relation have been evaded, at a pure phenomenological level, by assuming the existence of  "axion-like" particles with independent mass and couplings to the gauge sector (see \cite{Roncadelli:2017idg} for an overview). \\
A well-defined and complete gauge model where such interactions can be realized  is string theory, where several axion solutions are present, for many vacua and geometric compactifications \cite{Hui:2016ltb}. Moduli fields with flat potentials, of geometric origin, with either CP-even or CP-odd properties, which generate axion-like 
 ($a F\tilde{F}$) and dilaton ($\phi F F$) interations
abound in such constructions, and are generically motivated in low energy phenomenological descriptions. In supersymmetric theories, for instance, they are produced by Ferrara-Zumino supercurrents by the superconformal anomaly multiplet \cite{Coriano:2014gja} and can be conveniently discussed in specific extensions of the NMSSM with a Stueckelberg supermultiplet \cite{Coriano:2010ws,Coriano:2008aw,Coriano:2008xa}.\\
String axions have been studied in several earlier works
\cite{Witten,Antoniadis:2002qm, Coriano:2005js, Conlon,Svrcek,Choi,Arvanitaki,Choi2,Cicoli}.
However, it is of wide interest to provide field theoretical realizations for such particles, at least in the CP-odd case, within a well-defined gauge structure at lower energy, as discussed in \cite{Coriano:2005js}. This may justify axion-like fields in the spectrum of an ordinary gauge theory, while at the same time associate them to a fundamental principle such as a different mechanism of anomaly cancellation, beyond ordinary charge assignment of a chiral fermion spectrum, as in the case of the Standard Model.\\
 This was the approach of 
\cite{Coriano:2005js}, later investigated, independently from its string theory motivations \cite{Antoniadis:2002qm}, in bottom-up extensions of the Standard Model with one extra $U(1)$ \cite{Coriano:2006xh,Coriano:2008xa}. Such extensions incorporate a mechanism of gauge anomaly cancellations with the inclusion of Wess-Zumino terms. Such a method, which can be viewed as the field theoretical realization of the analogous Green-Schwarz mechanism of string theory, 
involves the gauging of an axionic symmetry and its breaking via an "extra potential" with the inclusion of the (real) Stueckelberg field - at least in the non supersymmetric case - beside the Higgs system, in the scalar potential. Such extra potential was motivated just on symmetry grounds in \cite{Coriano:2005js}.

A direct analysis of this term in \cite{Coriano:2010py,Coriano:2010ws} entertained the possibility that such periodic potential ($V_p(H_i,b)$), function of the Stuckelberg pseudoscalar $b$ and the Higgs fields $H_i$, could be generated at any phase transition, not just at the QCD one, and be periodic in a component of $b$, denoted by $\chi$, called in  
\cite{Coriano:2005js} "the Axi-Higgs". The Stuckelberg field is in general rewritten in terms of the physical, gauge invariant component, $\chi$, plus Goldstone fields which disappear in the unitary gauge. \\
The mixing of CP-odd 
phases of the Higgs sector and the Stueckelberg, which takes from $b(x)$ to $\chi(x)$ in this scenario \cite{Coriano:2010py}, parallels the mixing in the DFSZ (or weak) axion solution \cite{Dine:1981rt, Zhitnitsky:1980tq} of the strong CP problem.
Such phase-dependent potentials are naturally supported by a non-abelian gauge theory, and their size related to either weak or strong instantons \cite{Coriano:2010ws}. \\
Supersymmetric extensions of such 
constructions via a Stueckelberg supermultiplet \cite{Coriano:2010ws, Coriano:2008aw,Armillis:2007tb} 
and studies of the relic density of $\chi$ have also been investigated.\\
 In the case of supersymmetric models both scalars and pseudoscalar couplings can be consistently described, due to the presence of a dilaton \cite{Coriano:2010ws}. In both cases the fundamental principle upon which such models are built is the cancellation of gauge anomalies with Stueckelberg fields or Stueckelberg supermultiplets. 
 
 In this work we are going to present an extension of such formulations, in the non supersymmetric case, to an $E_6$ Grand Unified Theory (GUT), suitably generalized, which opens the way to lifting the previous bottom-up scenarios up to the Planck scale. This allows the possibility of setting the Stueckelberg mass $M_S$, previously arranged around the TeV scale \cite{Coriano:2007xg,Coriano:2009zh,Armillis:2008vp} larger or equal to the GUT scale $M_S\geq M_{GUT}$, with the mass of the anomalous $U(1)$ gauge 
 boson of the same order. The only extra assumption, in the scenario that we discuss, is the existence of an anomalous $U(1)$ symmetry, that we call $U(1)_X$. This can be naturally present at the Planck scale, with $M_S\sim M_{Planck}$ and realized in the Stueckelberg form, by the dual of an antisymmetric rank-2 tensor $C_{\mu\nu}$ coupled to the field strength $B_{\mu\nu}$ of the anomalous $U(1)$, via a $C\wedge$ B interaction. Within superstring theory models \cite{Dine:1987xk,Atick:1987gy,Dine:1987gj }, due to the appearance of Fayet-Iliopoulos terms, such constructions are naturally present. They carry significant implications for the breaking of supersymmetry by gaugino condensation \cite{Binetruy:1996uv} and the generation of neutrino mass textures, which have been extensively investigated in the past \cite{Ibanez:1994ig,Binetruy:1996xk,Binetruy:1996cb,Irges:1998ax,Coriano:2006rg}. We build a plausible and realistic model with the features that it is string-inspired
and is a GUT  which has both an invisible axion and an ultra-light
Stueckelberg pseudoscalar. The latter will be incorporated by a Green-Schwarz mechanism
to cancel the chiral anomaly of the $U(1)_X$ gauge symmetry. The key requirement for the generation of a physical axion is the generation of the phase-dependent potential mentioned above at the GUT scale $M_{GUT}$, which induces Higgs-axion mixing. The outcome of this construction is the appearance of a pseudoscalar which has been invoked in order to solve several astrophysical shortcomings in the search for dark matter, such as the missing satellite  problem and others \cite{Hui:2016ltb} which point towards an ultralight axion in the ${10^{-22}}-10^{-20}$ eV mass range. 
 
 \section{The Stueckelberg phase} 
 We consider a gauge symmetry of the form $E_6\times U(1)_X$, where the gauge boson $B^\mu$ is in the Stueckelberg phase. We define $B_{\alpha}$ as the gauge field of $U(1)_X$
and $B_{\alpha\beta} \equiv \partial_{\alpha}B_{\beta} - \partial_{\beta}B_{\alpha}$ as the corresponding
field strength and with $g_B$ its gauge coupling. The $U(1)_X$ carries an anomalous coupling to the fermion spectrum, that will be specified below. As in \cite{Ghilencea:2002da} we assume the presence of a term of the form $M_S C^{\alpha\beta}B^{\rho\sigma}\epsilon_{\alpha\beta\rho\sigma}$ in the bosonic sector, motivated by low energy string theory. 
The 1-particle irreducible (1PI) effective Lagrangian of the theory at 1-loop level takes the form 
\begin{equation}
{\cal L}={\cal L}_{E_6}+ {\cal L}_{Stueck} + {\cal L}_{anom} + {\cal L}_{WZ},
\end{equation}
in terms of the gauge contribution of $E_6$ (${\cal L}_{E_6}$), the Stueckelberg term ${\cal L}_{Stueck}$, the anomalous 3-point functions  ${\cal L}_{anom}$, generated by the anomalous fermion couplings to the $U(1)_X$ gauge boson, and the Wess-Zumino counterterm (WZ) ${\cal L}_{WZ}$. 
The kinetic term for the 2-form $C_{\alpha\beta}$ is the 3-form
\begin{equation}
H_{\alpha\beta\gamma} \equiv  \partial_{\alpha}C_{\beta\gamma} + \partial_{\gamma}C_{\alpha\beta} + \partial_{\beta}C_{\gamma\alpha},
\label{3form}
\end{equation}
whereupon the Stueckelberg addition to the $E_6$ gauge Lagrangian
\begin{equation}
{\cal L}_{E_6}=-\frac{1}{4} F^{(E_6)\, \mu\nu}F^{(E_6)}_ {\mu\nu}, 
\end{equation}
at tree level, is
\begin{equation}
{\cal L}_{Stueck}  = - \frac{1}{12} H_{\alpha\beta\gamma} H^{\alpha\beta\gamma}
- \frac{1}{4} B_{\alpha\beta}B^{\alpha\beta} + \frac{M_S}{4} \epsilon^{\alpha\beta\gamma\delta} C_{\alpha\beta} B_{\gamma\delta}. 
\end{equation}
\noindent We treat $C_{\alpha\beta}$ and $H_{\alpha\beta\gamma}$ as if independent in a first-order formalism
and introduce the Stueckelberg scalar $b(x)$ as a Lagrangian multiplier to re-write
\begin{equation}
{\cal L}_{Stueck} = - \frac{1}{12} H_{\alpha\beta\gamma} H^{\alpha\beta\gamma}
- \frac{1}{4} B_{\alpha\beta}B^{\alpha\beta} - \frac{M_S}{6} \epsilon^{\alpha\beta\gamma\delta}H_{\alpha\beta\gamma}B_{\delta} + \frac{1}{6} b(x) \epsilon^{\alpha\beta\gamma\delta}\partial_{\alpha}H_{\beta\gamma\delta}.
\end{equation}
\noindent
Integration by parts gives
\begin{equation}
H^{\alpha\beta\gamma} = - \epsilon^{\alpha\beta\gamma\delta} \left(M_S B_{\delta} - \partial_{\delta} b(x) \right)
\end{equation}
\noindent
which enables us to write the Stueckelberg part of the lagrangian as
\begin{equation}
{\cal L}_{Stueck} = - \frac{1}{4} B_{\alpha\beta} B^{\alpha\beta} - \frac{1}{2} (M_S B_{\alpha} - \partial_{\alpha} b(x))^2.
\end{equation}
\noindent
In this final form, $M_S$ is the mass of the Stueckelberg gauge boson associated with $U(1)_X$ which we can 
assume, as pointed out above, to be larger than the GUT scale, possibly coinciding with the Planck scale, guaranteeing the decoupling of the axion around $M_{GUT}$, due to the gravitational suppression of the WZ counterterms.  
The WZ contribution is the combination of two terms 
\begin{equation}
{\cal L}_{WZ}= c_1 \frac{b(x)}{M_S}F^{(E_6)\, \mu\nu}F^{(E_6)\, \rho\sigma}\epsilon_{\mu\nu\rho\sigma} + 
c_2  \frac{b(x)}{M_S}B_{\mu\nu}B_{\rho\sigma}\epsilon^{\mu\nu\rho\sigma} 
\label{WZ}
\end{equation}
needed for the cancellation of the $U(1)_X E_6 E_6$  and $U(1)_X^3$ anomalies respectively, with $c_1$ and $c_2$ numerical constants fixed by the charge assignments of the model. 
\section{Gravitational Effects}

\noindent
It is well known that couplings to gravity can endanger the stability of invisible
axion solutions to the strong CP problem. Three contemporaneous and independent
papers 
\cite{Barr:1992qq,Kamionkowski:1992mf,Holman:1992us} pointed out this issue already in 1992. In order
to avoid the problem, couplings to gravity must be suppressed for 
operators up to dimension ten, {\it i.e.} Planck scale operators with 
dimensions between five and nine, which might otherwise be added to a renormalizable
theory, must be avoided.
\noindent
One of the three papers \cite{Holman:1992us} made a start to constructing a GUT which enforced the needed suppression of dangerous operators. In the present
article we explore further such a theory especially with respect to predicting not only
a QCD axion but also an ultra-light Stueckelberg pseudoscalar candidate for dark matter.\\
The method to disallow operators of dimensions between five and nine is to choose an
appropriate GUT gauge group and irreps for the matter fields.
In \cite{Holman:1992us} the proposal was to use an ${\cal N}=1$
supersymmetric theory with GUT gauge group $E(6) \times U(1)_X$ and fields in irreps
${\bf 27}_{+1}$, ${\bf 27}_{-1}$ and ${\bf \overline{351}}_{0}$, where the subscripts designate
the corresponding $X$-charges. 
The only term allowed in a renormalizable superpotential is ${\bf 27}_{+1} {\bf 27}_{-1} {\bf 351}_{0}$
and the resultant theory possesses a global $U(1)_{PQ}$ symmetry with PQ charges $+1$ for
the ${\bf 27}$'s and $-2$ for the $\overline{\bf 351}$.  The key point is then that the lowest dimensional
terms in the effective lagrangian which are $E(6) \times U(1)_X$ gauge invariant
and break global $U(1)_{PQ}$ are ${\bf 27}^6$, ${\bf 351}^6$ and $({\bf 27}.{\bf 27}.{\bf 351})^2$
all of which correspond to dimension 10 as required. This leads to an invisible
axion in the mass range between 1 $\mu$eV and  $1$ meV and an acceptable solution of the strong CP problem.\\
\noindent
Instead, we shall build a {\it non}-supersymmetric GUT model based also on a gauge
group $E(6) \times U(1)_X$ which contains an invisible axion which
solves the strong CP problem similarly
to the supersymmetric model outlined in \cite{Holman:1992us}, but now contains
in addition an
ultra-light Stueckelberg psudoscalar which can play the role of dark matter as in \cite{Hui:2016ltb}.
\noindent
The $U(1)_X$ starts out as an anomalous $U(1)_X$ and the anomaly is then cancelled by the inclusion of additional Wess-Zumino terms as in \cite{Coriano:2005js, Coriano:2007xg}. These give rise to the ultra-light interactions between the Stueckelberg pseudoscalar and the gauge fields of the model via the two typical $(b(x)/M_S) F\tilde{F}$ anomaly counterterms. Such interactions are naturally inherited by $\chi$  after a rotation from the gauge to the mass eigenstates in the scalar CP-odd sector \cite{Coriano:2005js,Coriano:2008xa,Coriano:2009zh}.
\bigskip

\section {The Model}

One of the earliest phenomenologically promising superstring theories \cite{Gross:1984dd} 
was the heterotic string based on the gauge group
$E(8) \times  E(8)$. In this special case, compactification of six spatial dimensions on a Calabi-Yau manifold
gives rise naturally \cite{Candelas:1985en} to an $E(6)$ gauge GUT theory. Several alternative
GUT groups including $SU(5)$ and $SO(10)$ have since been derived from alternative
superstring theories with various compactifications.
In the case of $E(6)$ each fermion family appears in a ${\bf 27}$ irreducible representation of $E(6)$.
By itself and unadorned, such a gauge theory has no chiral anomaly.
$E(6)$ has the advantage that a simple $E(6)$ GUT with an invisible axion
is already known and it is this pre-existing model which is the skeleton on which
we append the new ultra-light particle.
\noindent
We shall be further inspired by superstring theory via the Green-Schwarz mechanism
\cite{Green:1984sg} which
will underly the cancellation of an additional anomaly and give rise to our ultra-light
Stueckelberg pseudoscalar. The point is that an anomalous abelian $U(1)$ has
a non-vanishing one-loop triangle diagram with Stueckelberg gauge bosons
at each of the three vertices. The Green-Schwarz mechanism cancels this
{\it one-loop} diagram by a {\it tree} diagram involving a pseudoscalar which has vertices
coupling it to both one and two gauge bosons. This additional particle will
provide the needed ultra-light dark matter.\\
\noindent
We build a model theory with gauge symmetry $E(6) \times U(1)_X$
where the original field of \cite{Frampton:1981ik} is anomalous with respect to $U(1)_X$ and therefore requires for consistency the introduction \cite{Coriano:2005js} of a Stueckelberg gauge
field $B_{\mu}$ and a Stueckelberg pseudoscalar field $b(x)$ as will be discussed in detail below.\\
The three chiral familes will be assigned under $E(6)\times U(1)_X$ respectively to
\begin{equation}
{\bf 27}_{X_1}  ~~~~~ {\bf 27}_{X_2} ~~~~~ {\bf 27}_{X_3},
\label{chiral}
\end{equation}
in which the charges $X_i$ ($i=1,2,3$) are to be determined.\\
Absence of the $U(1)_X^3$ and $E_6\times U(1)_X^2$ anomalies then would require that
\begin{equation}
\sum_{i=1}^{1=3} X_i^3 = 0,     \qquad\qquad \sum_{i=1}^3 X_i=0,
\label{Xcharges}
\end{equation}
which need to be violated in order to compensate with a Wess-Zumino term for the restoration of the gauge symmetry of the action.\\
The scalar sector of the model contains two ${\bf 351}_{X_i}$ $(i=1,2)$ irreducible representations,
where the $U(1)_X$ charges $X_i$ need to be determined. We note that ${\bf 351}$ is the {\it antisymmetric} part of the Kronecker product ${\bf 27} \otimes {\bf 27}$ where ${\bf 27}$ is the defining representation of $E(6)$. \\
Denoting a ${\bf 351}_X$ conveniently by the 2-form $A_{\mu\nu} = - A_{\nu\mu}$ with $\mu,\nu = 1$ to $27$, the 
most general renormalizable potential  in ${\cal L}_{E_6}$ is expressed in terms of two 2-forms $A^{(1)}_{\mu\nu}$ and $A^{(2)}_{\mu\nu}$ of $U(1)_X$ of charges $x_1$ and $x_2$ respectively. \\
Let us denote the ${\bf 27}_{X_i}$ of Eq.(\ref{chiral}) by $\Psi_{\mu}$ with $\mu = 1$ to $27$. Then
 the full lagrangian including the potential $V$, which will be specified below, has an invariance under the global symmetry 
 \begin{equation}
 A^{(1)}_{\mu\nu} \rightarrow e^{i\theta} A^{(1)}_{\mu\nu}~~~~~A^{(2)}_{\mu\nu} \rightarrow e^{i\theta} A^{(2)}_{\mu\nu} ~~~~ \Psi_{\mu} \rightarrow e^{- (\frac{1}{2}i\theta)} \Psi_{\mu},
 \label{PQ}
 \end{equation}
 which is identifiable as a Peccei-Quinn symmetry which is broken at the GUT scale when $E(6)$
 is broken to $SU(5)$ as discussed in \cite{Frampton:1981ik}, giving rise to an invisible axion
 in the mass range from approximately $10^{-12} \textrm{eV}$ to $10^{-2} \textrm{eV}$ and providing a solution to the strong CP problem.  
 \noindent
 We couple $A^{(1)}_{\mu\nu}$ to the fermion families $({\bf 27})_{X_i}$ $i=1,2,3$, choosing
 in Eq. (\ref{chiral}), {\it e.g.} $X_1=X_2=X_3=+1$, with the $X$-charge of $A^{(1)}$ fixed to $X=-2$. The second scalar representation $A^{(2)}$ is decoupled from the fermions, with an 
 $X-$charge for $A^{(2)}$ which is arbitrary and taken for simplicity to be $X=+2$. The coupling of the fermion spectrum to $U(1)_X$ is anomalous and consistency of the theory requires anomaly cancellation by addition of the Stueckelberg field $b(x)$, by the inclusion of (\ref{WZ}). The potential is then expressed in terms of three $E_6\times U(1)_X$ invariant components, 
 
\begin{equation}
V= V_1+  V_2 + V_p,
\end{equation}
where
\begin{equation} 
V_1= F(A^{(1)},A^{(1)})  \qquad  V_2= F(A^{(2)},A^{(2)}),
\end{equation}
with $V_1$ and $V_2$ denoting the contributions of $({\bf 351})_{-2}$ and $({\bf 351})_{+2}$, expressed in terms of 
the function \cite{Frampton:1981ik}
\begin{eqnarray}
 F(A^{(i)},A^{(j)}) & = &\left. M_{GUT}^2 A^{(i)}_{\mu\nu} \bar{A^{(j)}}^{\mu\nu} +h_1 ~(A^{(i)}_{\mu\nu} \bar{A^{(j)}}^{\mu\nu})^2  +h_2  ~ A^{(i)}_{\mu\nu} \bar{A}^{\nu\sigma} A^{(i)}_{\sigma\tau}\bar{A}^{\tau\mu} \right.\nonumber \\
&&\left.\qquad \qquad +\,h_3 ~  d^{\mu\nu \lambda} d_{\xi\eta\lambda} A^{(i)}_{\mu\sigma}A^{(i)}_{\nu\tau} \bar{A^{(j)}}^{\xi\sigma} \bar{A^{(j)}}^{\eta\tau} \right. \nonumber \\
 & &\left.  \qquad \qquad +\, h_4 ~ d^{\mu\nu\alpha}d^{\sigma\tau\beta}d_{\xi\eta\alpha} d_{\lambda\rho\beta} A^{(i)}_{\mu\sigma}A^{(i)}_{\nu\tau} \bar{A^{(j)}}^{\xi\lambda} \bar{A^{(j)}}^{\eta\rho}\right. \nonumber \\
 & &\left. \qquad \qquad + \,h_5 ~ d^{\mu\nu\alpha} d^{\sigma\beta\gamma} d_{\xi\eta\beta} d_{\lambda\alpha\gamma} A^{(i)}_{\mu\sigma}A^{(i)}_{\nu\tau} \bar{A^{(j)}}^{\xi\lambda} \bar{A^{(j)}}^{\eta\tau} \right.\nonumber \\
 & &\left. \qquad \qquad +\,h_6 ~ d^{\mu\nu\alpha}d^{\sigma\tau\beta} d_{\alpha\beta\gamma} d^{\gamma\zeta\xi} d_{\xi\eta\zeta} d_{\lambda\rho\chi} A^{(i)}_{\mu\sigma}\bar{A^{(j)}}^{\xi\lambda} A^{(i)}_{\nu\tau} \bar{A^{(j)}}^{\eta\rho}\right. ,
 \end{eqnarray}
 in which $d_{\alpha\beta\gamma}$ with $\alpha,\beta,\gamma = 1$ to $27$ is the $E(6)$
invariant tensor.\\
Due to the asymmetric X-charge assignment and the presence of two scalar sectors, then we are allowed to introduce the extra contribution, which is allowed by symmetry 
\begin{eqnarray}
V_p & = & M_{GUT}^2 A^{(1)}_{\mu\nu} \bar{A^{(2)}}^{\mu\nu}e^{- i 4\frac{b}{M_S}}   + e^{- i 8 \frac{b}{M_S}}\left[(h_1 ~(A^{(1)}_{\mu\nu} \bar{A^{(2)}}^{\mu\nu} )^2  + h_2  ~ A^{(1)}_{\mu\nu} \bar{A^{(2)}}^{\nu\sigma} A^{(1)}_{\sigma\tau}\bar{A^{(2)}}^{\tau\mu} \right.\nonumber \\
&&\left.\qquad \qquad +\,h_3 ~  d^{\mu\nu \lambda} d_{\xi\eta\lambda} A^{(1)}_{\mu\sigma}A^{(1)}_{\nu\tau} \bar{A^{(2)}}^{\xi\sigma} \bar{A^{(2)}}^{\eta\tau} \right. \nonumber \\
 & &\left.  \qquad \qquad +\, h_4 ~ d^{\mu\nu\alpha}d^{\sigma\tau\beta}d_{\xi\eta\alpha} d_{\lambda\rho\beta} A^{(1)}_{\mu\sigma}A^{(1)}_{\nu\tau} \bar{A^{(2)}}^{\xi\lambda} \bar{A^{(2)}}^{\eta\rho}\right. \nonumber \\
 & &\left. \qquad \qquad + \,h_5 ~ d^{\mu\nu\alpha} d^{\sigma\beta\gamma} d_{\xi\eta\beta} d_{\lambda\alpha\gamma} A^{(1)}_{\mu\sigma}A^{(1)}_{\nu\tau} \bar{A^{(2)}}^{\xi\lambda} \bar{A^{(2)}}^{\eta\tau} \right.\nonumber \\
 & &\left. \qquad \qquad +\,h_6 ~ d^{\mu\nu\alpha}d^{\sigma\tau\beta} d_{\alpha\beta\gamma} d^{\gamma\zeta\xi} d_{\xi\eta\zeta} d_{\lambda\rho\chi} A^{(1)}_{\mu\sigma}\bar{A^{(2)}}^{\xi\lambda} A^{(1)}_{\nu\tau} \bar{A^{(2)}}^{\eta\rho}\right] + h.c.
 \end{eqnarray}
and which becomes periodic at the GUT scale after symmetry breaking, similarly to the case considered in 
\cite{Coriano:2010py,Coriano:2010ws}. This potential, in our case, is expected to be of nonperturbative origin. In particular, the size of the 
 contributions in $V_p$, generated by instanton effects at the GUT scale, are expected to be exponentially suppressed.

 \subsection{The periodic potential}
The breaking of the $E_6\times U(1)_X$ symmetry at $M_{GUT}$ can follow different routes such as  
  $E(6) \supset SU(3)_C \times SU(3)_L \times SU(3)_H$, the ${\bf 351}^{'}$ or $E(6) \supset SU(5)$. In the first case 
\begin{eqnarray}
\label{351prime}
({\bf 351}) & = & (1, 3^*, 3) + (1, 3^*,6^*) + (1, 6, 3) + (3, 3, 1) + (3, 6^*, 1) + (3, 3, 8) + \nonumber \\
& & (3^*, 1, 3^*) + (3^*, 1, 6)
+ (3^*, 8, 3^*) + (6^*, 3, 1) + (6, 1, 3^*) + (8, 3^*, 3)
\label{351}
\end{eqnarray}
of which the colour singlets are only the 45 states for each of the two $({\bf 351})_{X_i}$
\begin{equation}
(1, 3^*, 3)_{X_i} \qquad   (1, 3^*, 6^*)_{X_i} \qquad (1, 6, 3)_{X_I}  ~~~  i=1,2.
\label{colorsinglets}
\end{equation}
Bearing in mind that, for $SU(3)_L \supset SU(2)_L$, $3^* = 2 +1$ and $6=3 + 2 + 1$ we see that
there are exactly nine colour-singlet $SU(2)_L$-doublets in the $({\bf 351}^{'})_{-2}$ and 9 in the 
$({\bf 351}^{'})_{+2}$, that we may denote as  $H^{(1)}_j$, $H^{(2)}_j$, with $j=1,2\ldots 9$, which appear in the periodic potential in the form 
\begin{eqnarray}
V_p &\sim&\sum_{j=1}^{12}\lambda_0 M_{\textrm {GUT}}^2 (H^{(1)\dagger }_j H^{(2)}_j e^{- 4 i g_B\frac{b}{M_S}})+
\sum_{j,k=1}^{12}\left[\lambda_1(H^{(1)\dagger}_j H^{(2)}_j e^{-i 4 g_B \frac{b}{ M_S}})^2+\lambda_2(H_i^{(1)\dagger}H_i)(H_i^{(1)\dagger}H^{(2)}_j e^{-i 4 g_B\frac{b}{M_S}})\right.\nonumber \\
&&\left. + \lambda_3(H_k^{(2)\dagger}H_k^{(2)})(H_j^{(1)\dagger}H_k^{(2)} e^{-i 4 g_B\frac{b}{M_S}}) \right] +\textrm{h.c.},
\end{eqnarray}
where we are neglecting all the other terms generated from the decomposition (\ref{351prime}) which will not contribute to the breaking. We assume that such potential, which is allowed by the gauge symmetry, is instanton generated at the GUT scale, with parameters $\lambda_i$'s which, as already mentioned, are exponentially suppressed.\\
For simplicity we will consider only a typical term in the expression above, involving two neutral components, generically denoted as $H^{(1)\, 0}$ and $H^{(2)\, 0}$, being all the remaining contributions similar. In this simplified case the axi-Higgs $\chi$ is generated by the mixing of the CP odd components of two neutral Higgses. We parameterize such components in polar form as

\begin{equation}
H^{(1)\, 0}=\frac{1}{\sqrt{2}}\left(\sqrt{2}v_1 + \rho_1^0(x) \right) e^{i\frac{F_1^0(x)}{\sqrt{2}v_1}}
\hspace{1cm}
H^{(2)\,0}=\frac{1}{\sqrt{2}}\left(\sqrt{2}v_2 + \rho_2^0(x) \right) e^{i\frac{F_2^0(x)}{\sqrt{2}v_2}},
\end{equation}
where we have introduced the two phases $F_1$ and $F_2$ of the two neutral Higgs fields and the two vevs $v_1$ and $v_2$,  which in our case are of $O(M_{\textrm{GUT}}).$
Equivalently, the potential can be described in terms of the imaginary components of $H^{(1)\,0}$ and $H^{(2)\,0}$. In this case, denoting with $q_1$ and $q_2$ the two $U(1)_X$ charges 
of the two Higgses, the diagonalization of this sector by an orthogonal matrix $O^\chi$ allows to rotate the St\"uckelberg field and the CP-odd phases of the two Higgs doublets into the mass eigenstates 
$(\chi, G^{\,0}_1, G^{\,0}_2)$. They correspond to a physical (gauge invariant) axion
$\chi$ and to two Goldstone modes $G^0_1$ and $G^{0}_2$. We indicate with $O^{\chi}$ the orthogonal matrix which allows to rotate them on the physical basis
\begin{equation}
\begin{pmatrix}
G_0^1 \cr
G_0^2 \cr
\chi
\end{pmatrix}
= O^\chi
\begin{pmatrix}
\textrm{Im}H^0_d \cr
\textrm{Im}H^0_u \cr
b
\end{pmatrix},
\end{equation}
with 
\begin{equation}
O^\chi=
\begin{pmatrix}
\frac{v_2}{v} & \frac{v_1}{v} & 0 \cr
-\frac{g_B (q_2-q_1)v_2 v_1^2}{v\sqrt{g_B^2 (q_2-q_1)^2 v_2^2 v_1^2+2 M_S^2 v^2}}&
\frac{g_B (q_2-q_1)v_2^2 v_1}{v\sqrt{g_B^2 (q_2-q_1)^2 v_2^2 v_1^2+2 M_S^2 v^2}} &
\frac{\sqrt{2}M_S v}{\sqrt{g_B^2 (q_2-q_1)^2 v_2^2 v_1^2+2 M_S^2 v^2}} \cr
\frac{\sqrt{2} M_S v_1}{\sqrt{g_B^2 (q_2-q_1)^2 v_1^2 v_2^2+2 M_S^2 v^2}}&
-\frac{\sqrt{2} M v_2}{\sqrt{g_B^2 (q_2-q_1)^2 v_1^2 v_2^2+2 M_S^2 v^2}} &
\frac{g_B (q_2-q_1) v_2 v_1}{\sqrt{g_B^2 (q_2-q_1)^2 v_1^2 v_2^2+2M_S^2 v^2}} 
\end{pmatrix}
\label{stella}
\end{equation}
$\chi$ inherits WZ interaction since $b$ can be related to the physical axion $\chi$ and to the Goldstone modes via this matrix 
\begin{equation}
b =  O_{13}^{\chi} G_0^1 + O_{23}^{\chi} G_0^2 + O_{33}^{\chi} \chi ,       
\label{rot12}
\end{equation}
Here $v=\sqrt{v_1^2+v_2^2}$ and the explicit expressions of the mass eigenstates, in this simplified case are

\begin{align}
G_0^1&=\frac{1}{\sqrt{v_1^2+v_2^2}}(v_1,v_2,0)\nonumber\\
G_0^2&=\frac{1}{\sqrt{g_B^2 (q_2-q_1)^2 v_2^2 v_1^2+2 M_S^2 \left(v_2^2+v_1^2\right)}}\left(-\frac{ g_B (q_2-q_1)v_2 v_1^2}{\sqrt{v_1^2+v_2^2}},\frac{g_B (q_2-q_1)v_2^2 v_1}{\sqrt{v_2^2+v_1^2}},\sqrt{2}M_S\sqrt{v_1^2+v_2^2}\right)\nonumber\\
\chi&=\frac{1}{\sqrt{g_B^2 (q_2-q_1)^2 v_1^2 v_2^2+2 M_S^2(v_2^2 + v_1^2)}}
\left(\sqrt{2} M_S v_1,-\sqrt{2} M_S v_2, g_B (q_2-q_1) v_2 v_1\right).
\end{align}

The potential $V_p$ is periodic with respect to the linear combination of fields
\begin{equation}
\theta(x)\equiv\frac{ 4 g_B }{M_S}b(x)-\frac{1}{\sqrt{2}v_1} F_1^0(x) +\frac{1}{\sqrt{2}v_2} F_2^0(x),
\end{equation}
and can be rotated to the physical basis in the CP odd sector with the same matrix 
$O^\chi$. It is proportional to the physical axion $\chi(x)$, modulo a dimensionful constant ($\sigma_\chi$)
\begin{equation}
\theta(x)\equiv \frac{\chi(x)}{\sigma_\chi},
\end{equation}
where we have defined
\begin{equation}
\sigma_\chi\equiv\frac{2  v_1 v_2 M_S}{\sqrt{ g_B^2  v_1^2 v_2^2 +2 M_S^2 (v_1^2+v_2^2)}}.
\end{equation}
Notice that $\sigma_\chi$, in our case, takes the role of $f_a$ of the PQ case, where the angle of 
misalignment is identified by the ratio $a/f_a$, with by $a$ the PQ axion. The two Goldstone modes generated by $O^\chi$, in two different linear combinations, are absorbed one by the anomalous gauge boson and the other by a neutral gauge boson generated by the breaking of the GUT symmetry.
Therefore, generalizing this procedure, the structure of $V_{p}$ after the breaking of the $E_6\times U(1)_X$ symmetry can be summarised in the form 

\begin{align}
V_p\sim&  v_1 v_2
\left(\lambda_2 v_2^2+\lambda_3 v_1^2+\overline{\lambda_0} M_{GUT}^2\right) \cos\left(\frac{\chi}{\sigma_\chi}\right) +  \lambda_1 v_1^2 v_2^2 \cos\left(2\frac{\chi}{\sigma_\chi}\right),
\label{extrap}
\end{align}
with a mass for the physical axion $\chi$  given by
\begin{equation}
m_{\chi}^2\sim\frac{2 v_1 v_2}{\sigma^2_\chi}\left(\bar{\lambda}_0 v_1^2 +\lambda_2 v_2^2 +\lambda_3 v_1^2+4 \lambda_1 v_1 v_2\right) 
\approx \lambda v^2
\label{axionmass}
\end{equation}
with $v_1\sim v_2\sim v\sim M_{GUT}$.
It is quite straightforward to provide estimates for $\sigma_\chi$ and $m_\chi$. Assuming that $M_S$, the Stueckelberg 
mass, is of the order of $M_{\textrm{Planck}}$ and that the breaking of the $E_6\times U(1)_X$ symmetry takes place at 
$M_{GUT}\sim 10^{15}$ GeV,  (e.g. $v_1\sim v_2\sim M_{GUT}$) then 
\begin{equation}
\sigma_{\chi}\sim M_{\textrm{GUT}} + {\cal O}(M_{\textrm{GUT}}^2/M_{\textrm{Planck}}^2),  \qquad m_\chi^2\sim \lambda_0 M_{\textrm{GUT}}^2,
\end{equation}
where all the $\lambda_i$'s in $V_p$ are of the same order.
Being $V_p$ generated by an instanton sector, then $\lambda_0\sim e^{-2 \pi/\alpha(M_{\textrm{GUT}})}$, with the value of the coupling  $4 \pi g_B^2 =\alpha_{\textrm{GUT}}$ fixed at the GUT scale. If we assume that  $1/33 \le \alpha_{GUT} \le 1/32$, then  $e^{-201}\sim 10^{-91}\le \lambda_0 \le e^{-205}\sim 10^{-88}$, and
the mass of the axion $\chi$ takes the approximate value 
\begin{equation}
      10^{-22}   \textrm{ eV} < m_{\chi} <  10^{-20} \textrm{ eV}, 
\end{equation}
which is in the allowed mass range for an ultralight axion, as discussed in recent analysis of the astrophysical constraints on this type of dark matter \cite{Hui:2016ltb}.
 \section{Discussion}

 \noindent
 The strong CP problem is one of the biggest difficulties for the standard model and
 the QCD axion provides the most elegant and simple known solution. However, the allowed mass range
 for this particle is $10^{-12}\textrm{ eV} \leq m_a \leq 10^{-3}\textrm{ eV}$. The lower bound
 arises from the requirement that the Peccei-Quinn breaking scale be below the Planck
 scale.
  \noindent
 The small scale ($\leq 10\, \textrm{kpc}$) anomalies of cold dark matter require an even lighter
 axion-like particle with $10^{-22}\,\textrm{eV} \leq m_a^{'} \leq10^{-20}\,\textrm{eV}$, clearly
 non-overlapping with the QCD axion mass range, and so suggests
 we require at least two different axion-like particles.
 \noindent
 Inspired by the Green-Schwarz mechanism, a second axion-like particle ($\chi$) occurs naturally in the present model from the cancellation of an anomalous $U(1)_X$ gauge symmetry via a Stueckelberg pseudoscalar which shifts under this symmetry, and therefore {\it both} these two axion-like states may exist in Nature
 and should be sought for by experiment. \\
 The QCD axion can be sought, as now, by terrestrial
 experiments while the ultra-light pseudoscalar probably can be inferred only from astronomical
 observations concerning dark matter and their comparison with numerical simulations.
 The importance
 of our new model is that it shows how one can have, in the same theory, {\it both} an axion which solves the strong CP problem
 {\it and} an ultra-light axion which provides a cold dark matter candidate. We hope to return with a more extended analysis of this scenario in some related work.

 \centerline{\bf Acknowledgements} 
This research was partially supported by INFN within (Iniziativa Specifica) QFT-HEP.
\noindent



\end{document}